\documentclass[12pt]{article}

\usepackage{amssymb}
\usepackage{amsfonts}

\begin{document}

\title{Quantum feedback control in quantum photosynthesis}

\author{S.V. Kozyrev\footnote{Steklov Mathematical Institute of Russian Academy of Sciences, Moscow, Russia}, A.N. Pechen$^{*}$}

\maketitle

\begin{abstract}
A model of charge separation in quantum photosynthesis as a model of quantum feedback control in a system of interacting excitons and vibrons is introduced. Quantum feedback in this approach describes the Landau--Zener transition with decoherence. The model explains irreversibility in the process of charge separation for quantum photosynthesis --- direct transitions for this quantum control model will have probabilities close to one and reverse transitions will have probabilities close to zero. This can be considered as a model of quantum ratchet. Also this model explains coincidence of energy of the vibron paired to the transition and Bohr frequency of the transition.
\end{abstract}

\section{Introduction}

Quantum effects in photosynthetic systems attract a lot of attention \cite{Engel}. In particular irreversibility of charge separation in quantum photosynthesis was discussed \cite{Irreversible}. One of key observations is the presence of vibrons in photosynthetic systems --- vibrations of nuclear degrees of freedom of chromophores with lifetime of order of several picoseconds \cite{Novoderezhkin2015}. The following phenomenon was observed --- usually energy of the vibron is close to the Bohr frequency (difference of energies of the levels) for electronic states which interact with the vibron, moreover one vibron can interact with several transitions with the same Bohr frequency  \cite{Novoderezhkin2015}, \cite{Novoderezhkin2017}.

Exciton--vibron interaction works as follows. Interaction of electrons at chromophores with light generates excitons. Excitons generate vibrons as vibrations of nuclei in the field of exciton according to the Franck--Condon principle (in the semiclassical form) --- landscape of potential energy for nuclei (which corresponds to Coulomb interaction) changes abruptly  when exciton is excited, at this new energy landscape positions of nuclei are non-equilibrium and nuclei begin to oscillate around new minima of potential energy. Transitions between potential energy surfaces for electronic states happen in vicinity of avoided crossing point (where potential energy surfaces are close) using the Landau--Zener mechanism. It was observed \cite{Novoderezhkin2019} that lifetime of electronic coherences (important for the Landau--Zener mechanism) is similar to the time of transition between the energy levels for the process of charge separation in photosynthesis. Therefore we actually have to consider Landau--Zener transitions with friction which takes in account the decoherence.

Different approaches to description of vibrons in quantum photosynthesis were discussed, in particular consideration of excitons and vibrons as collective eigenstates  of the Hamiltonian \cite{Novoderezhkin2015}, \cite{Novoderezhkin2017}, related to this approach master equations in the polaron frame \cite{Kolli2011}, \cite{Kolli2012}, \cite{Cao}, master equations with non-secular terms \cite{Chin2013} (vibrons as coherences of vibrational degrees of freedom generated by interaction with populations of electronic states should be described by non-secular terms in master equations).  The considered in this paper model of charge separation explains both irreversibility of charge separation and coincidence of energy of the vibron paired to the transition and Bohr frequency of the transition.

According to \cite{Novoderezhkin2019} the process of charge separation in bacterial reaction center utilizes interaction of electronic states with two vibrons with energies 115 and 35 cm$^{-1}$. Vibron with energy 115 cm$^{-1}$ is generated when exciton state is created and oscillates along the reaction coordinate for transition between exciton state and charge separation state causing a transition to the charge separation state. Vibron with energy 35 cm$^{-1}$ is generated after this transition and prevents recombination of the charge transfer state providing directionality of the process of charge separation.

In the present paper we construct a model of charge separation in quantum photosynthesis based on quantum feedback control. Quantum feedback in this model corresponds to decoherence of electronic degrees of freedom during the Landau--Zener transition (described by quantum measurement procedure, or collapse of electronic wave function) and makes this transition directed. Landau--Zener theory with dissipation was considered in \cite{Frauenfelder}, \cite{Arceci} (without discussion of possible relation to quantum feedback). Quantum control attracts attention in relation to quantum computations and quantum technologies, see in particular \cite{Barchielli}, \cite{Milburn}, \cite{Pechen1}, \cite{Pechen2}, \cite{Pechen3}, \cite{JohnGough}, \cite{Belavkin}. We describe irreversible transition by a model of quantum ratchet which is related to quantum Maxwell's daemon. A model of (classical) Maxwell's daemon was considered by Smoluchowski \cite{Smoluchowski} and discussed by Feynman in his lectures \cite{Feynman}, the daemon was based on ratchet which performs separation of molecules. Charge separation in the photosynthetic reaction center in our model utilizes quantum ratchet which performs irreversible transitions between potential energy surfaces of electronic states using oscillations of vibrons. The quantum ratchet contains two vibrons --- vibron ${\bf q}^{(1)}(t)$ in formula below moves electronic state $|1\rangle$ to the avoided crossing point and vibron ${\bf q}^{(2)}(t)$ in this formula repels state $|2\rangle$ from the avoided crossing point after the transition. These two vibrons (with energies 115 and 35 cm$^{-1}$ correspondingly \cite{Novoderezhkin2019}) combined control irreversible charge separation transition (the vibrons can be called the right and the left hands of quantum Maxwell's daemon). The quantum control model under consideration
$$
\frac{d}{dt}\psi(t)=-iH(\psi,t)\psi(t),\quad \psi(t)=\pmatrix{\langle 1 | \psi(t)\rangle\cr \langle 2 | \psi(t)\rangle},
$$
\begin{equation}\label{intro1}
H(\psi,t)=\pmatrix{{\bf h}_1\cdot{\bf q}^{(1)}(t)|\langle 1 | \psi(t)\rangle|^2 & J\cr J & {\bf h}_2\cdot\left({\bf q}^{(1)}(t)+{\bf q}^{(2)}(t)\right)|\langle 2 | \psi(t)\rangle|^2}
\end{equation}
contains cubic nonlinearity of the kind of nonlinear Schroedinger equation. Here time dependent vectors ${\bf q}^{(1)}(t)$, ${\bf q}^{(2)}(t)$ describe vibrons --- special modes of vibrations of chromophores and $\psi(t)$ describes electronic degrees of freedom (described by two potential energy surfaces for exciton $|1\rangle$ and charge separation state $|2\rangle$).

Experimental observations of photosynthetic complexes reveal many vibrons paired to different transitions between electronic states \cite{Novoderezhkin2017}, \cite{Detrapping}. Charge separation as a principal process is paired with two vibrons which allows to make this transition directed. Some other transitions are paired to a single vibron ${\bf q}(t)$ with corresponding quantum feedback control equation
$$
\frac{d}{dt}\psi(t)=-iH(\psi,t)\psi(t),\quad \psi(t)=\pmatrix{\langle 1 | \psi(t)\rangle\cr \langle 2 | \psi(t)\rangle},
$$
\begin{equation}\label{intro2}
H(\psi,t)=\pmatrix{{\bf h}_1\cdot{\bf q}(t)|\langle 1 | \psi(t)\rangle|^2 & J\cr J & {\bf h}_2\cdot{\bf q}(t)|\langle 2 | \psi(t)\rangle|^2}.
\end{equation}

A model of this kind can be proposed for a process of avoiding of trapping in quantum photosynthesis described in \cite{Detrapping}. One could consider quantum feedback control as important mechanism for efficient operation of photosynthetic systems.

\section{Vibrons and quantum feedback control}

In the present section we construct a model of quantum feedback control for the quantum photosynthetic system discussed in \cite{Novoderezhkin2019}.

\medskip

\noindent{\bf Landau--Zener formula}.
The model by Landau and Zener investigates the dynamics
$$
\frac{d}{dt}\psi(t)=-iH(t)\psi(t)
$$
with time dependent generator acting in the basis of states $|1\rangle$, $|2\rangle$ (called diabatic energy levels, eigenlevels of the time dependent Hamiltonian $H(t)$ are called adiabatic levels)
$$
H(t)=\pmatrix{ u t & J\cr J & v t},\quad |1\rangle=\pmatrix{1 \cr 0},\quad |2\rangle=\pmatrix{0 \cr 1}.
$$

Then the probability of transition between diabatic states at large times will be equal to $1-P$ where
$$
P=e^{-2\pi\gamma},\quad \gamma=\frac{J^2}{|u-v|}.
$$

Actually the transition occurs in vicinity of the avoided crossing point where matrix elements at the diagonal of the generator are equal to zero, i.e. $t=0$. Value of $1-P$ is large for large interaction $J$ or low difference $|u-v|$ of velocities of passage through the avoided crossing point at different slopes of the transition.

\medskip

\noindent
{\bf Landau--Zener transition with friction as a model of quantum feedback}.
Landau--Zener transitions with decoherence (measurement assisted transitions) were considered in \cite{Pechen3}. This approach is not valid if projection occurs at the avoided crossing. It might seem that probability of collapse of wave function in vicinity of avoided crossing is rather low but there exist important regimes where this probability can not be neglected.

Time of transitions $|1\rangle\rightarrow|2\rangle$ of charge separation in \cite{Novoderezhkin2019} is approximately 100 fs which is close to lifetime of electronic coherences.
Without decoherence the electronic wave function ''feels both sides of the transition due to quantum nonlocality'' which gives the contribution $|u-v|$ in the Landau--Zener formula $P=e^{-\frac{2\pi J^2}{|u-v|}}$ and probabilities of direct and reverse transitions are equal, but if this wave function is collapsed in the avoided crossing point between the slopes $u$ and $v$ probabilities of direct and reverse transitions will become different (with $|u|$ and $|v|$ instead of $|u-v|$). It is natural to choose the avoided crossing as the point of collapse of wave function for the regime when the vibron is slow in the avoided crossing --- vicinity of the avoided crossing is small but the system spends there a lot of time (see the discussion below).

Discussion of Landau \cite{Landau} involves quasiclassical arguments, the transition probability (denoted by $P$ above) is given by integration over complex time (where $t_1$ is real and $t_2$ is the branch point where two eigenvalues of the Hamiltonian are equal)
$$
\exp\left(-\frac{2}{\hbar}{\rm Im}\int_{t_1}^{t_2}(E_2(t)-E_1(t))dt\right),
$$ 
the difference of eigenvalues arise from complex contour integral where the integral at the left side of the contour contains $E_1(t)$ and the integral at the right side of the contour contains $E_2(t)$.

If the wave function is collapsed at the branch point $t_2$ only half of integral remains and we obtain for transition probability
$$
\exp\left(\frac{2}{\hbar}{\rm Im}\int_{t_1}^{t_2}E_1(t)dt\right)
$$
as stated above.

Decoherence in exciton--vibron interaction was discussed by some kind of master equation (see in particular \cite{Egorova2004}, \cite{Novoderezhkin2016}).
Nonlinearity in the approach of the present paper can be considered as effect of approximation of self--interaction as in the method of self--consistent field.

\medskip

\noindent{\bf Exciton--phonon Hamiltonian}. The Hamiltonian for interaction between excitons and phonons has the form \cite{YangFleming}
$$
H=H^{\rm el}+H^{\rm Coul}+H^{\rm ph}+H^{\rm el-ph}+H^{\rm reorg}
$$
$$
H^{\rm el}=\sum_{n}\varepsilon_n|n\rangle\langle n|,\quad H^{\rm reorg}=\sum_{nk}|n\rangle\langle n|\int\frac{h_{nk}^2(\xi)}{\omega_k(\xi)}d\xi,
$$
$$
H^{\rm ph}=\sum_k\int \omega_k(\xi)c^*_k(\xi)c_k(\xi)d\xi,\quad H^{\rm el-ph}=\sum_{nk}|n\rangle\langle n|\int h_{nk}(\xi)\left(c^*_k(\xi)+c_k(\xi)\right)d\xi,
$$
$$
H^{\rm Coul}=\frac{1}{2}\sum_{mn,m\ne n}J_{mn}|m\rangle\langle n|,\quad J_{nm}=J_{mn}^*.
$$

Here $|n\rangle$ are electronic states (excitons), $c^*_k$ are creation operators for the field of phonons, values $J_{mn}$ describe Coulomb interaction between excitons.

\medskip

\noindent{\bf Vibrons and quantum feedback control}.
To describe the process of charge separation we need three electronic states: $|0\rangle$ --- the ground state, $|1\rangle$ --- exciton, $|2\rangle$ --- charge separation state, $\varepsilon_1>\varepsilon_2>\varepsilon_0$. The above Hamiltonian contains the transition term between $|1\rangle$, $|2\rangle$ (where $J$ is real)
$$
J\left(|1\rangle\langle 2|+|2\rangle\langle 1|\right).
$$

We will understand vibrons classically (as vector--functions of time), i.e. to describe interaction between electrons and vibrons we will substitute the quantum operator $c_k+c_k^*$ by classical value $q_k$ in the above Hamiltonian $H^{\rm el-ph}$ (this corresponds to the semiclassical Franck--Condon principle). Then interaction of the vibron with the pair of electronic states will be described by the formula
$$
\sum_{n=1,2;k}|n\rangle\langle n|h_{nk}q_k(t)=|1\rangle\langle 1|{\bf h}_1\cdot{\bf q}(t) +|2\rangle\langle 2|{\bf h}_2\cdot{\bf q}(t).
$$
Here vibron is a vector ${\bf q}(t)=(q_k(t))$ in the scalar products with vectors ${\bf h}_{1,2}$ of interactions with electronic states.

Moreover we have to take in account that vibrons are switched on by transitions between electronic states: excitation of the exciton $|1\rangle$ generates the vibron ${\bf q}^{(1)}(t)$ and transition to the charge transfer state $|2\rangle$ generates also the second vibron ${\bf q}^{(2)}(t)$. Therefore we obtain feedback  --- the quantum dynamics controls itself --- generation of vibrons is controlled by populations of corresponding electronic states, vibrons initiate transitions between electronic states. To describe this feedback we will multiply projectors $|n\rangle\langle n|$, $n=1,2$ to electronic states in the above formula by populations $\rho_{nn}(t)$ of these states. This procedure describes phenomenologically the effect of measurement i.e. decoherence of electronic wave function due to interaction with the environment, moreover the measurement occurs at the avoided crossing point.

For the dynamics of exciton--vibron interaction we get the generator of Landau--Zener dynamics with friction for a quantum feedback control model
$$
H(\rho,t)=\pmatrix{{\bf h}_1\cdot{\bf q}^{(1)}(t)\rho_{11}(t) & J\cr J & {\bf h}_2\cdot\left({\bf q}^{(1)}(t)+{\bf q}^{(2)}(t)\right)\rho_{22}(t)},
$$
where all vibrons as functions of time should be continuous and should have continuous first derivatives (a vibron after excitation starts moving from some initial position and zero velocity).

For the case of pure quantum states under consideration
$$
\rho_{11}(t)=|\langle 1 | \psi(t)\rangle|^2,\quad \rho_{22}(t)=|\langle 2 | \psi(t)\rangle|^2,
$$
where $\psi$ is a wave function of the electron. We get the (phenomenological) quantum feedback control equation
\begin{equation}\label{Daemon}
\frac{d}{dt}\psi(t)=-iH(\psi,t)\psi(t),\quad \psi(t)=\pmatrix{\langle 1 | \psi(t)\rangle\cr \langle 2 | \psi(t)\rangle},
\end{equation}
\begin{equation}\label{Daemon1}
H(\psi,t)=\pmatrix{{\bf h}\cdot{\bf q}^{(1)}(t)|\langle 1 | \psi(t)\rangle|^2 & J\cr J & -{\bf h}\cdot\left({\bf q}^{(1)}(t)+{\bf q}^{(2)}(t)\right)|\langle 2 | \psi(t)\rangle|^2},
\end{equation}
where we choose ${\bf h}_1=-{\bf h}_2={\bf h}$.

Squared moduli of quantum amplitudes at the diagonal of the generator $H(\psi,t)$ describe collapse of electronic wave function in the avoided crossing point for Landau--Zener transition with friction. Quantum feedback model (\ref{Daemon}), (\ref{Daemon1}) has cubic nonlinearity as for nonlinear Schroedinger equation. We claim that this model allows to obtain different probabilities of transitions for direct $|1\rangle\rightarrow|2\rangle$ and reverse $|2\rangle\rightarrow|1\rangle$ transitions (hence it describes quantum ratchet) although the Landau--Zener formula predicts equal probabilities for direct and reverse transitions.
Let us perform a rough estimate of probabilities of transitions for quantum ratchet (\ref{Daemon}), (\ref{Daemon1}).

\medskip

Let us consider the direct move of the quantum ratchet  --- transition $|1\rangle\rightarrow|2\rangle$ where $\langle 1 | \psi(t)\rangle=1$ and $\langle 2 | \psi(t)\rangle=0$. Then the generator of the transition takes the form
$$
\pmatrix{{\bf h}\cdot{\bf q}^{(1)}(t) & J\cr J & 0}.
$$

Avoided crossing point for the direct move of the quantum ratchet is defined by the condition
$$
{\bf h}\cdot{\bf q}^{(1)}(t)=0.
$$

Since vibronic oscillations are at least of order of magnitude slower compared to electronic ones these oscillations can be linearized in vicinity of the avoided crossing point. Then probability of transition $|1\rangle\rightarrow |2\rangle$ between diabatic states for one passage of the vibron through the avoided crossing point can be approximated by the Landau--Zener formula (with quantum feedback i.e. electronic wave function is collapsed in the avoided crossing point): this probability equals $1-P$ where
$$
P=e^{-2\pi\gamma},\quad \gamma=\frac{J^2}{|{\bf h}\cdot{\bf v}|},
$$
and ${\bf v}$ is velocity of the vibron in the moment of passage of the avoided crossing point.  By deceleration of the vibron at the moment of the passage it is possible to make the transition probability close to one $1-P\approx 1$.

Let us consider the reverse move of the quantum ratchet  --- transition $|2\rangle\rightarrow|1\rangle$ where $\langle 1 | \psi(t)\rangle=0$ and $\langle 2 | \psi(t)\rangle=1$. The generator of the transition (we again take in account the quantum feedback) is
$$
\pmatrix{0 & J\cr J & -{\bf h}\cdot\left({\bf q}^{(1)}(t)+{\bf q}^{(2)}(t)\right)}.
$$

For the reverse move of the quantum ratchet the avoided crossing point satisfies
$$
{\bf h}\cdot\left({\bf q}^{(1)}(t)+{\bf q}^{(2)}(t)\right)=0.
$$

Instead of single vibron ${\bf q}^{(1)}(t)$ (for the direct move) this transition is driven by the sum of vibrons ${\bf q}^{(1)}(t)$ and ${\bf q}^{(2)}(t)$. Hence it might happen that the above equation has no solutions and the transition probability is close to zero $1-P\approx 0$. In general the problem of quantum control for the model (\ref{Daemon}), (\ref{Daemon1}) of the quantum ratchet can be formulated as follows: find parameters for vibrons  ${\bf q}^{(1)}(t)$ and ${\bf q}^{(2)}(t)$ to make transition probability for the direct move $|1\rangle\rightarrow |2\rangle$ of the ratchet close to one and transition probability for the reverse move $|2\rangle\rightarrow |1\rangle$ of the ratchet make close to zero.

\section{A model of the quantum ratchet}

Let us consider a simple model for vibrons and show that the picture of operation of the quantum ratchet described in the previous section indeed takes place.

The ansatz for the first vibron is as follows
\begin{equation}\label{Vibron1}
{\bf q}^{(1)}(t)={\bf q}_0 + \theta(t-t_1){\bf v}_1\left(\cos \omega_1 (t-t_1) -1\right),
\end{equation}
where $\theta$ is the step function (equal to zero for negative argument and to one for non-negative argument) i.e. when the vibron is switched on at time moment $t_1$ (when exciton is excited) it begins to oscillate along vector ${\bf v}_1$ with frequency $\omega_1$, with zero initial velocity starting from the initial position ${\bf q}_0$ (which corresponds to the stationary nuclear coordinates in the ground state $|0\rangle$ of the electron).

For the second vibron the ansatz has the form
\begin{equation}\label{Vibron2}
{\bf q}^{(2)}(t)=\theta(t-t_2){\bf v}_2\left(\cos \omega_2 (t-t_2)-1\right),
\end{equation}
i.e. the vibron is switched on at time moment $t_2$ and oscillates with frequency $\omega_2$ along vector ${\bf v}_2$, initial displacement and initial velocity are equal to zero.

Equation of the avoided crossing point for the direct move of the quantum ratchet is
\begin{equation}\label{AW1}
{\bf h}\cdot\left[{\bf q}_0 + {\bf v}_1\left(\cos \omega_1 (t-t_1)-1\right)\right]=0.
\end{equation}

This equation can be satisfied if $2{\bf h}\cdot{\bf v}_1\ge {\bf h}\cdot {\bf q}_0$ (i.e. if the amplitude of oscillations of the vibron is large enough). Let us assume for simplicity that scalar products by ${\bf h}$ of vectors ${\bf v}_1$, ${\bf q}_0$ are positive.

To elevate transition probability one has to decrease velocity of the vibron (\ref{Vibron1}) in the moment of passage through the avoided crossing point (\ref{AW1}) i.e. to make $2{\bf h}\cdot{\bf v}_1\approx {\bf h}\cdot {\bf q}_0$. In this case velocity of the vibron in the avoided crossing point will be close to zero. To amplify transitions (to make transition probability for the direct move of the ratchet close to one) one has to decrease the amplitude of the vibron almost to minimal value which still allows existence of solution of equation (\ref{AW1}) of the crossing point (we call this regime {\em optimal}).

The energy difference between adiabatic levels (eigenlevels of the time dependent Hamiltonian) in the initial point $t=t_1$ of the vibron is close to ${\bf h}\cdot {\bf q}_0$ (at $t=t_1$ adiabatic levels are close to diabatic levels $|1\rangle$, $|2\rangle$) and in the avoided crossing point it equals almost zero (since $J$ is small), i.e. in the optimal regime energy of the vibron matches the energy difference between the levels. Approximate coincidence of energy of the vibron and energy difference of levels coupled to this vibron was widely discussed in the literature, in particular in \cite{Novoderezhkin2015}, \cite{Novoderezhkin2017}, \cite{Novoderezhkin2019}.

Equation of the avoided crossing point for the reverse move of the quantum ratchet has the form
\begin{equation}\label{AW2}
{\bf h}\cdot\left[{\bf q}_0 + {\bf v}_1\left(\cos \omega_1 (t-t_1)-1\right)+{\bf v}_2\left(\cos \omega_2 (t-t_2)-1\right)\right]=0.
\end{equation}

Solvability of this equation can be easily broken, especially in the optimal regime for the direct move of the ratchet. In the optimal regime the sum of the first two terms in (\ref{AW2}) will hardly cross zero. Let the third term in (\ref{AW2}) gives a non-negative contribution in the scalar product with ${\bf h}$ (i.e. the scalar product ${\bf h}\cdot {\bf v}_2$ should be negative) then this contribution will be close to zero at time moments when the sum of the first two terms is positive. In this regime equation of the avoided crossing point (\ref{AW2}) for the reverse move of the quantum ratchet will not have solutions and reverse transitions will be effectively forbidden.

Time parameter $t_1$ in (\ref{Vibron1}) is the initial condition for (\ref{Daemon}), (\ref{Daemon1}) (a moment of creation of exciton). Parameter $t_2$ in (\ref{Vibron2}) is a moment of transition between diabatic states $|1\rangle$, $|2\rangle$ i.e. this time moment is related to a solution of (\ref{Daemon}). For simplicity let us make $t_2$ equal to the time moment when population of the level $|2\rangle$ in (\ref{Daemon}), (\ref{Daemon1}) will exceed some threshold (if we start from the level $|1\rangle$ and only the first vibron ${\bf q}^{(1)}(t)$ is present in the generator). This definition is analogous to the definition of classical probability theory of the moment when the Brownian motion reaches a certain boundary (in general, random variables of this kind are called Markov moments).

\medskip

\noindent
{\bf Remark}. Discussion of irreversibility above uses the presence of two different vibrons in generator (\ref{Daemon1}). In  (\ref{intro1}), (\ref{intro2}) we also had different vectors ${\bf h}_1$, ${\bf h}_2$ in generators (i.e. not necessarily ${\bf h}_1=-{\bf h}_2$). Non-parallelism of ${\bf h}_1$ and ${\bf h}_2$ can also be utilized to make the corresponding transition irreversible even for the case of single vibron: it might happen that equations of avoided crossing point of the type (\ref{AW1}) with ${\bf h}_1$ and ${\bf h}_2$ instead of ${\bf h}$ will behave as follows: the first equation is solvable and the second does not have solutions (if ${\bf q}_0$ and ${\bf v}_1$ are non-parallel).

\medskip

\noindent
{\bf Remark}. Discussed in the present paper model of charge separation has the following feature: in order to make charge transfer faster vibron (\ref{Vibron1}) operates in the ''optimal'' regime i.e. the amplitude of the vibron has the minimal value which allows to satisfy equation (\ref{AW1}) of the avoided crossing point. It was mentioned above that in this regime energy of the vibron matches the energy difference between the levels. Also in this regime the vibron is ''fragile'' --- perturbations of parameters of the vibron related to mutations can break the solvability of (\ref{AW1}) and block charge transfer. In \cite{Novoderezhkin2019} mutations in photosynthetic reaction center were investigated, with the following observations: the YM210W mutation which causes small changes in geometry and potential in the reaction center modifies the operation of the vibron with energy 115 cm$^{-1}$ and slows the excitation transfer by two orders of magnitude. One can discuss this observation as a proof that vibron with energy 115 cm$^{-1}$ works in the ''optimal'' regime (velocity of the vibron in the avoided crossing point is slow). It is interesting that the GM203L mutation which removes the second vibron (with energy 35 cm$^{-1}$) slows the excitations transfer by only one order of magnitude \cite{Novoderezhkin2019}.

\medskip

\noindent
{\bf Summary}. We considered a model of charge separation transition in quantum photosynthesis based on quantum feedback control equation. This quantum feedback control model is explained as follows: for Landau--Zener transition with friction and transitions close to adiabatic (the vibron is slow at avoided crossing) collapse of electronic wave function should occur at the avoided crossing point. This can be modeled by quantum feedback which allows to make probabilities of direct and reverse transitions different (the transition becomes directed). Taking in account excitation of two vibrons it is possible to make this transition maximally irreversible (i.e. to make probability of direct transition maximal and of reverse transition minimal). To make probability of the direct transition maximal energy of the vibron should be close to the Bohr frequency of the transition (which is observed in experiments).

\medskip

\noindent
{\bf Acknowledgments.}
This work was funded by the Ministry of Science and Higher Education of the Russian Federation (grant number 075-15-2020-788). The authors are grateful to I.V.Volovich and A.S.Trushechkin for discussions.

\end{document}